\documentclass{article}

\usepackage{amssymb}
\usepackage{amsmath}
\usepackage[dvips]{graphicx}

\begin{document}

\noindent \textbf{Estimation of the effects of Y and Zn atoms on the elastic properties of Mg solid solution}
\newline
\newline
\newline

\noindent Vassiliki Katsika-Tsigourakou*\newline

\noindent \textit{Department of Solid State Physics, Faculty of Physics, University of Athens, }

\noindent \textit{Panepistimiopolis, 157 84 Zografos, Greece}\newline

\noindent \textbf{Abstract}\newline

\noindent Mg-based alloys have recently attracted major interest in view of their potential applications in the aerospace, aircraft and automotive industries. Here, we show that the effects of Y and Zn atoms on their compressibility can be reliably estimated by a simple thermodynamical model deduced by means of first-principles calculations based on density functional theory that just appeared.\newline

\noindent \textbf{PACS (2008)}: 61.72.J, 62.20.de, 66.30.Fq, 61.72.Bb


\noindent \_\_\_\_\_\_\_\_\_\_\_\_\_\_\_\_\_\_\_\_\_\_\_\_\_\_\_\_\_\_\_\_\_\_\_\_\_\_\_

\noindent *E-mail vkatsik@phys.uoa.gr

\newpage

\noindent \textbf{1. Introduction }\newline

\noindent Due to the high importance in modern technology, the effects of solid solution strengthening in Mg alloys has been recently investigated (e.g., see Refs [1-3]) using various kinds of solute atoms. After the observation [4] that the hardness of Mg-Zn alloys increases upon increasing the Zn content, additional investigations [5-7] revealed the strengthening effect of Zn in Mg-Rare Earths (RE) solid solutions. The latter solutions have attracted major attention in view of their high specific strength at both room and elevated temperatures and of their excellent creep resistance [1, 3, 8]. In particular, Y is considered to be one of the most important RE elements to improve mechanical properties of Mg-alloys at high temperatures [9].

\noindent Very recently [10] first-principles plane-wave pseudo-potential calculations based on density functional theory (DFT) appeared which studied the effects of Y and Zn atoms on elastic properties of Mg solid solutions. The bulk modulus $B$, shear modulus, Young's modulus and Poisson ratio were also derived [10] by Voigt-Reuss approximation, indicating a strong dependence of the elastic modulus on concentration of the solute atoms. Here, focusing solely on the bulk modulus (or its inverse, i.e., the compressibility $\kappa $, $\kappa ={1\mathord{\left/ {\vphantom {1 B}} \right. \kern-\nulldelimiterspace} B} $), we shall show that values comparable to those obtained by the aforementioned first-principles calculations can be determined on the basis of a thermodynamical model [13, 14] (for a recent review, see Ref. [15]). Note that such a procedure has already provided successful results for the compressibilities of ionic solid solutions, either of NaCl- or CsCl-type structure (see Refs [16] and [17] respectively), as well as for the transition-metal carbides and nitrides alloys Zr${}_{x}$Nb${}_{1-x}$C and Zr${}_{x}$Nb${}_{1-x}$N [18].\newline

\noindent \textbf{2. The procedure to estimate the compressibility of a solid solution in terms of the compressibilities  of the end members }\newline

\noindent Let us denote $V_{1} $ and $V_{2} $ the corresponding molar volumes, i.e. $V_{1} =N\upsilon _{1} $ and $V_{2} =N\upsilon _{2} $ (where $N$ stands for Avogadro's number) of two pure end members 1 and 2 respectively. A ``defect volume'' [19, 20] $\upsilon ^{d} $ is defined as the variation of the volume $V_{1} $, if one ``molecule'' of type $(1)$ is replaced by one ``molecule'' of type $(2)$. It is then evident that the addition of one ``molecule'' of type $(2)$ to a crystal containing $N$ ``molecules'' of type $(1)$ will increase its volume by $\upsilon _{}^{d} +\upsilon _{1} $. Assuming that $\upsilon ^{d} $ is independent of composition, the volume $V_{N+n} $ of a crystal containing $N$ ``molecules'' of type $(1)$ and  $n$ ``molecules'' of type $(2)$ can be written as: 
\begin{equation} \label{GrindEQ__1_} 
    V_{N+n} =[1+({n\mathord{\left/ {\vphantom {n N}} \right. \kern-\nulldelimiterspace} N} )]V_{1} +n\upsilon ^{d}                                           
\end{equation} 
         The compressibility $\kappa $ of this solid solution can be obtained by differentiating Eq. (1) with respect to pressure which gives [19]:            
\begin{equation} \label{GrindEQ__2_} 
          \kappa V_{N+n} =\kappa _{1} V_{1} +\left({n\mathord{\left/ {\vphantom {n N}} \right. \kern-\nulldelimiterspace} N} \right)\left[\kappa ^{d} N\upsilon ^{d} +\kappa _{1} V_{1} \right]
\end{equation} 
 where $\kappa ^{d} $ denotes the compressibility of the volume $\upsilon ^{d} $, defined as
\begin{equation} \label{GrindEQ__3_} 
\kappa ^{d} \equiv -({1\mathord{\left/ {\vphantom {1 \upsilon ^{d} )\times ({d\upsilon ^{d} \mathord{\left/ {\vphantom {d\upsilon ^{d}  dP)_{T} }} \right. \kern-\nulldelimiterspace} dP)_{T} } }} \right. \kern-\nulldelimiterspace} \upsilon ^{d} )\times ({d\upsilon ^{d} \mathord{\left/ {\vphantom {d\upsilon ^{d}  dP)_{T} }} \right. \kern-\nulldelimiterspace} dP)_{T} } } .                                                                 
\end{equation} 
To a first  approximation, the ``defect--volume'' $\upsilon ^{d} $ can be estimated from:
\begin{equation} \label{GrindEQ__4_} 
\upsilon ^{d} ={(V_{2} -V_{1} )\mathord{\left/ {\vphantom {(V_{2} -V_{1} ) N}} \right. \kern-\nulldelimiterspace} N}  
\end{equation} 
Thus, if $V_{N+n} $ can be determined in general from Eq. \eqref{GrindEQ__1_} (or approximately from  Eq. (4)), the compressibility $\kappa $ can be found from Eq. \eqref{GrindEQ__2_} in case that a reliable estimate of  $\kappa ^{d} $ can be made.

\noindent Along this direction, we employ the  thermodynamical model, termed $cB\Omega $ model, for the formation and migration of  the defects in solids [13-15]. This model has been checked for defect processes in a variety of solids [19] as well as in cases of activation of defects in complex ionic materials where upon gradually increasing the pressure ($P$) a critical pressure is reached at which an electric signal is emitted well before fracture thus providing an explanation for the detection of signals before major earthquakes [21, 22]. 

\noindent According to the aforementioned thermodynamical model, the defect Gibbs energy $g^{i} $ is interconnected with the bulk properties of the solid through the relation $g^{i} =c^{i} B\Omega $ where $B$ stands, as mentioned, for the isothermal bulk modulus $(=1/\kappa )$, $\Omega $ the mean volume per atom and $c^{i} $ is dimensionless quantity. (The superscript $i$ refers to the defect process under consideration, e.g. defect formation, defect migration and self-diffusion activation). By differentiating this relation in respect to pressure$P$, we find the defect volume $\upsilon ^{i} $ $[=({dg^{i} \mathord{\left/ {\vphantom {dg^{i}  dP)_{T} ]}} \right. \kern-\nulldelimiterspace} dP)_{T} ]} $. The compressibility $\kappa ^{d,i} $   defined  by   $\kappa ^{d,i} $$[\equiv -({d\ell n\upsilon ^{i} \mathord{\left/ {\vphantom {d\ell n\upsilon ^{i}  dP)_{T} ]}} \right. \kern-\nulldelimiterspace} dP)_{T} ]} $,  is given   by [19, 23]:
\begin{equation} \label{GrindEQ__5_} 
\kappa ^{d,i} =({1\mathord{\left/ {\vphantom {1 B)-{({d^{2} B\mathord{\left/ {\vphantom {d^{2} B dP^{2} )}} \right. \kern-\nulldelimiterspace} dP^{2} )} \mathord{\left/ {\vphantom {({d^{2} B\mathord{\left/ {\vphantom {d^{2} B dP^{2} )}} \right. \kern-\nulldelimiterspace} dP^{2} )}  [({dB\mathord{\left/ {\vphantom {dB dP)_{T} -1]}} \right. \kern-\nulldelimiterspace} dP)_{T} -1]} }} \right. \kern-\nulldelimiterspace} [({dB\mathord{\left/ {\vphantom {dB dP)_{T} -1]}} \right. \kern-\nulldelimiterspace} dP)_{T} -1]} } }} \right. \kern-\nulldelimiterspace} B)-{({d^{2} B\mathord{\left/ {\vphantom {d^{2} B dP^{2} )}} \right. \kern-\nulldelimiterspace} dP^{2} )} \mathord{\left/ {\vphantom {({d^{2} B\mathord{\left/ {\vphantom {d^{2} B dP^{2} )}} \right. \kern-\nulldelimiterspace} dP^{2} )}  [({dB\mathord{\left/ {\vphantom {dB dP)_{T} -1]}} \right. \kern-\nulldelimiterspace} dP)_{T} -1]} }} \right. \kern-\nulldelimiterspace} [({dB\mathord{\left/ {\vphantom {dB dP)_{T} -1]}} \right. \kern-\nulldelimiterspace} dP)_{T} -1]} } }  
\end{equation} 
We now assume that the validity of Eq. \eqref{GrindEQ__5_} holds also for the compressibility $\kappa ^{d} $ involved in Eq. \eqref{GrindEQ__2_}, i.e.,
\begin{equation} \label{GrindEQ__6_} 
                     \kappa ^{d} =\kappa _{1} -{({d^{2} B_{1} \mathord{\left/ {\vphantom {d^{2} B_{1}  dP^{2} )}} \right. \kern-\nulldelimiterspace} dP^{2} )} \mathord{\left/ {\vphantom {({d^{2} B_{1} \mathord{\left/ {\vphantom {d^{2} B_{1}  dP^{2} )}} \right. \kern-\nulldelimiterspace} dP^{2} )}  [({dB_{1} \mathord{\left/ {\vphantom {dB_{1}  dP)_{T} -1]}} \right. \kern-\nulldelimiterspace} dP)_{T} -1]} }} \right. \kern-\nulldelimiterspace} [({dB_{1} \mathord{\left/ {\vphantom {dB_{1}  dP)_{T} -1]}} \right. \kern-\nulldelimiterspace} dP)_{T} -1]} } 
\end{equation} 
            where the subscript ``1'' in the quantities at the right side denotes that they refer to the pure end member \eqref{GrindEQ__1_}. Since, in general, the quantities ${dB_{1} \mathord{\left/ {\vphantom {dB_{1}  dP}} \right. \kern-\nulldelimiterspace} dP} $ and ${d^{2} B_{1} \mathord{\left/ {\vphantom {d^{2} B_{1}  dP^{2} }} \right. \kern-\nulldelimiterspace} dP^{2} } $, can be roughly estimated from the modified Born model, Eq. \eqref{GrindEQ__6_} can provide an estimate of  $\kappa ^{d} $ and therefrom-on the basis of Eq. \eqref{GrindEQ__2_} the compressibility $\kappa $ of the solid solution can be found at various concentrations. This procedure has been applied in the former cases [16-18]. In the present case, however, we do not apply this procedure, because it is not certain that the modified Born model can satisfactorily describe the Mg solid solutions under consideration. Thus, we apply the following alternative procedure.

\noindent An inspection of Fig. 3(b) of Ref. [10] verifies that the volume $V$ varies almost linearly with respect to different concentrations of Y and Zn, thus agreeing with the essence of Eq. \eqref{GrindEQ__1_}. Hence, to a first approximation we can write 
\begin{equation} \label{GrindEQ__7_} 
V=(1-x)V_{1} +xV_{2}  
\end{equation} 
the differentiation of which in respect to pressure leads to:
\begin{equation} \label{GrindEQ__8_} 
B=B_{1} \left[\frac{1+x\left(\frac{V_{2} }{V_{1} } -1\right)}{1+x\left(\frac{B_{1} V_{2} }{B_{2} V_{1} } -1\right)} \right] 
\end{equation} 
This relation enables the estimation of the bulk modulus of the solid solution, for various concentrations, in terms of the bulk moduli $B_{1} $ and $B_{2} $ of the pure end members. The application of Eq. \eqref{GrindEQ__8_} leads to values that strikingly coincide with those computed in Ref. [10] (see their figure 5(a)) on the basis of first-principle calculations.

\noindent It should be emphasized that the success of the thermodynamical model [13, 14, 19] adopted here, has been also checked very recently in several tens of other mixed systems (see Ref. [24] along with its improvement in Ref. [25]).\newline

\noindent \textbf{3. Conclusion }\newline

\noindent Using a thermodynamical model which interconnects the point defect parameters in solids in terms of the bulk properties, we obtained a relation, i.e. Eq. \eqref{GrindEQ__8_}, which allows the estimation of the bulk modulus of solid solutions in terms of the bulk moduli of the end members. This gives results that strikingly coincide with those deduced very recently for Mg-based alloys by means of first-principle calculations based on density functional theory.

\newpage

\textbf{References}\newline

\noindent  [1] L. Gao, R. S. Chen, E. H. Han, J. Alloys Compd. 472, 234 (2009)

\noindent  [2] H. K. Lim, D. H. Kim, J. Y. Lee, W. T. Kim, J. Alloys Compd. 468, 308

(2009)

\noindent  [3] J. Zhang, J. Wang, X. Qiu, D. Zhang, Z. Tian, X. Niu, D. Tang, J. Meng, ,

J.  Alloys Compd. 464, 556 (2008)

\noindent  [4] D. Y. Maeng, T. S. Kim, J. H. Lee, S. J. Hong, S. K. Seo, B. S. Chun, Scr.

Mater. 43, 385 (2000)

\noindent  [5] X. B. Liu, R. S. Chen, E. H. Han,  J. Alloys Compd. 465, 232 (2008)

\noindent  [6] T. Homma, T. Ohkubo, S. Kamado, K. Hono, Acta Mater. 55, 4137 (2007)

\noindent  [7] Y. Gao, Q. Wang, J. Gu, Y. Zhao, Y. Tong, Mater. Sci. Eng. A 459, 117 

(2007) 

\noindent  [8] L. Gao, R. S. Chen, E. H. Han,  J. Alloys Compd. 481, 379 (2009)

\noindent  [9] M. Suzuki, T. Kimura, J. Koikie, K. Maruyama, Mater. Sci. Eng. 387, 706

 (2004)

\noindent [10] F. Yang, T. W. Fan, J. Wu, B. Y. Tang, L. M. Peng, W. J. Ding, Phys.  

Status  Solidi B 248, 2809 (2011) 

\noindent [11] P. Hohenberg, W. Kohn, Phys. Rev. 136, B864 (1964)

\noindent [12] W. Kohn, L. Sham, Phys. Rev. 137, A1697 (1965)

\noindent [13] P. Varotsos, Phys. Rev. B 13, 938 (1976)~;~P. A. Varotsos, J. Physique
 
(France)  Lettr. 38\textbf{,}  L455 (1977); Phys. Status Solidi B 90\textbf{,} 339 (1978)~;

 Phys. Status  Solidi B 100\textbf{,}  K133 (1980)

\noindent [14]  P. Varotsos, K. Alexopoulos, Phys.Rev\textit{. }B 15\textbf{,} 4111 (1977)~;           

15\textbf{,} 2348 (1977)~; 21\textbf{,} 4898 (1980)~; 24\textbf{,} 904 (1981)~; 30, 7305 (1984) 

\noindent [15] P. Varotsos, J. Appl. Phys. 101, 123503 (2007)

\noindent [16] V. Katsika-Tsigourakou, A. Vassilikou-Dova, J. Appl. Phys. 103,

083552 (2008)

\noindent [17] V. Katsika-Tsigourakou, Pramana- journal of Physics, 77, 689 (2011)

\noindent [18] V. Katsika-Tsigourakou, Cent. Eur. J. Phys. 9, 1309 (2011)

\noindent [19] P. Varotsos, K. Alexopoulos, Thermodynamics of Point Defects   

and Their Relation With the Bulk Properties, (North-Holland, 

Amsterdam, 1986)

\noindent [20] P. Varotsos, Phys. Status Solidi B 99\textbf{,} K93 (1980); J. Phys. Chem. 

Solids 42\textbf{,} 405 (1981); P. Varotsos, K. Alexopoulos, J. Phys. 

Chem. Solids 41\textbf{, }1291 (1980)

\noindent [21] P. Varotsos, K. Alexopoulos, Tectonophysics 110, 73 (1984)~; 

110, 99 (1984)~; P. Varotsos, M. Lazaridou, Tectonophysics

188, 321 (1991)

\noindent [22] P. Varotsos, K. Alexopoulos, K. Nomicos, M. Lazaridou, 

 Nature\textit{ }(London) 322\textbf{,} 120 (1986)~; Tectonophysics 152, 193 

(1988)~; P. A. Varotsos, N. Sarlis, E. S. Skordas, Phys. Rev. E  

66\textbf{,} 011902 (2002); 67\textbf{,} 021109 (2003); 68\textbf{,} 031106 (2003)

\noindent [23] P. Varotsos, W. Ludwig, K. Alexopoulos, Phys. Rev. B 18\textbf{, }2683 

(1978)~; P. Varotsos, W. Ludwig, J. Phys. C: Solid State  11\textbf{,} L305 

(1978); P. Varotsos, W. Ludwig, C. Falter, J. Phys. C: Solid State 

11\textbf{,} L311 (1978)\textbf{}

\noindent [24] O. Coreno-Alonso, J. Coreno-Alonso, J. Intermetallics 22, 142 

(2012)

\noindent [25] E. S. Skordas, J.Intermetallics doi:10.1016/j.intermet.2012.02.004

\end{document}